\documentclass{ws-ijmpe}

\begin{document}

\markboth{M.~Isse et al.}{
Jet-fluid string formation and decay
in high-energy heavy-ion collisions
}

\catchline{}{}{}{}{}

\title{
Jet-fluid string formation and decay
in high-energy heavy-ion collisions
}

\author{M.~Isse}
\address{
Department of Physics, Graduate School of Science,
Osaka University,\\
Toyonaka 560-0043, Japan\\
isse@kern.phys.sci.osaka-u.ac.jp }

\author{T.~Hirano\footnote{Present address:
Department of Physics, Graduate School of Science,
University of Tokyo, Tokyo 113-0033, Japan}}
\address{
Institute of Physics, Graduate School of Arts and Sciences,
University of Tokyo,\\
Komaba, Tokyo 153-8902, Japan
}
\author{R.~Mizukawa, A.~Ohnishi, K.~Yoshino}
\address{
Department of Physics, Faculty of Science,
Hokkaido University,\\
Sapporo 060-0810, Japan
}
\author{Y.~Nara}
\address{
Institut f\"ur Theoretische Physik,
Johann Wolfgang Goethe-Universit\"at,\\
60438 Frankfurt am Main, Germany
}

\maketitle

\begin{history}
\end{history}

\begin{abstract}
We propose a new hadronization mechanism,
jet-fluid string (JFS) formation and decay,
to understand observables in  intermediate to
high-$p_{T}$ regions comprehensively.
In the JFS model,
hard partons produced in jet lose their energy 
in traversing the QGP fluid,
which is described by fully three-dimensional hydrodynamic simulations.
When a jet parton escapes from the QGP fluid,
it picks up a partner parton from a fluid and forms a color singlet string,
then it decays to hadrons.
We find that high-$p_T$ $v_2$ values in JFS
are about two times larger than in the independent fragmentation model.
\end{abstract}

\section{Introduction}

Models based on the color strings have been highly successful to
describe high energy hadronic collisions.
A hadron-string cascade picture in high energy heavy-ion collisions
works well at AGS\cite{JAM} and SPS\cite{Isse} energies,
and it also describes some low-$p_T$ observables
such as $p_T$-distribution and $\eta$-distribution\cite{HINOY2005,Sahu}
at RHIC energies.
However, hadron-string cascade models underestimate the elliptic flow $v_2$
at RHIC energies,
therefore we need partonic pressure in the early stage.
 
%
\begin{figure}[b]
\begin{center}
\includegraphics[width=6cm,clip]{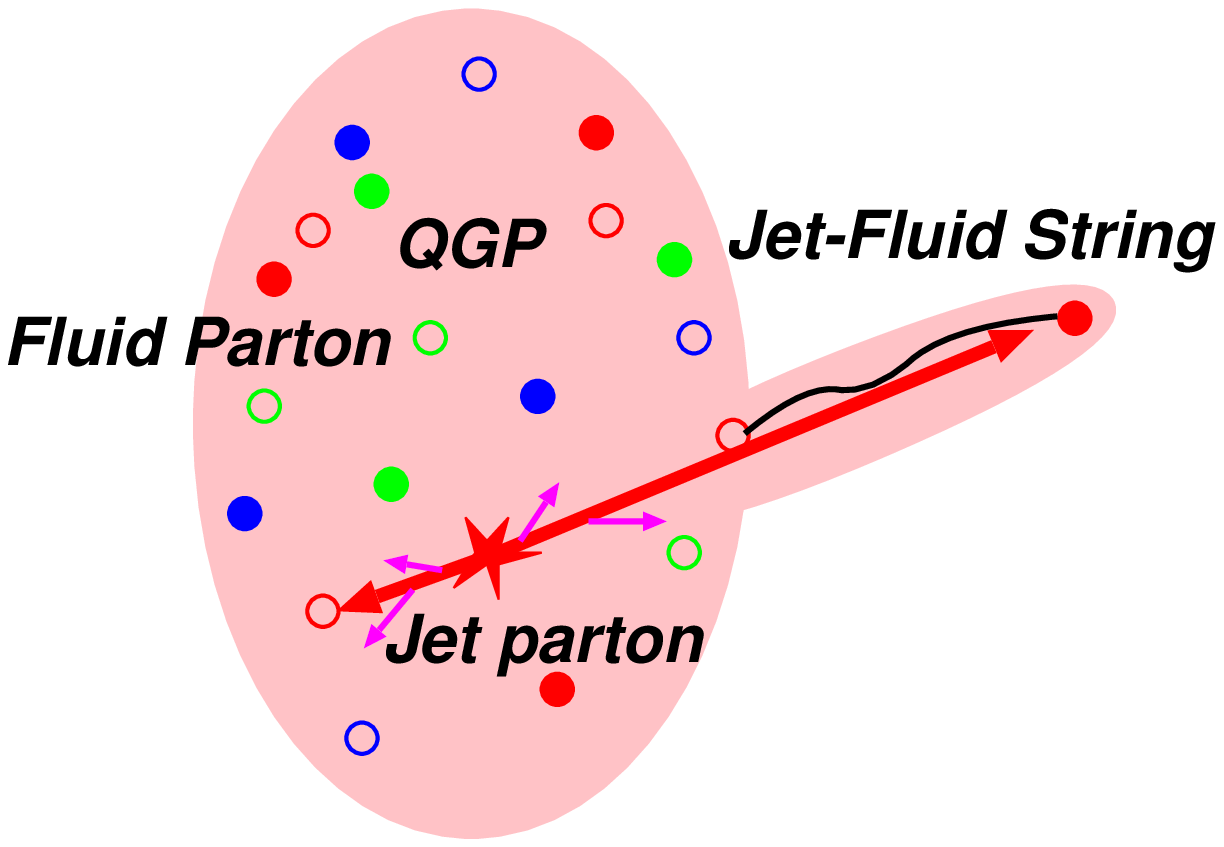}%
~\includegraphics[width=6cm,clip]{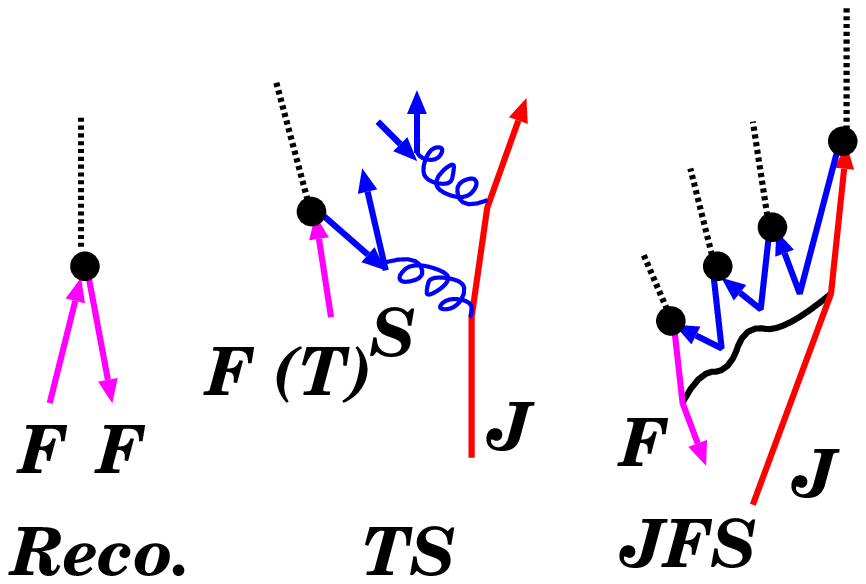}
\caption{
Left panel:
A sketch of JFS model.
Produced jet parton traverses in the evolving fluid with energy loss.
When it reaches to the QGP phase boundary, it picks up a fluid parton
and forms a string
and decays into hadrons.
Right panel:
Schematic view of simple Recombination,\protect\cite{Duke}
Thermal-Shower recombination,\protect\cite{Oregon}
and JFS fragmentation (See text).
}
\label{Fig:JFSform}
\end{center}
\end{figure} 
%
There are several observations explained as a
signal of quark-gluon plasma (QGP) formation at RHIC.
A large elliptic flow of bulk matter,
suppression of hadron yield at high-$p_T$,
and quark number scaling of elliptic flow at intermediate-$p_T$,
are explained as consequences of
hydrodynamical evolution in the early stage,\cite{HG2005}
parton energy loss in deconfined matter,\cite{e-loss}
and parton recombination,\cite{Duke} respectively.
All of these explanations support the formation of a QGP.
While these three pictures --- hydrodynamics, fragmentation, recombination ---
have succeeded in explaining many data at RHIC,
there are some problems left unsolved.
One of the problems is that
elliptic flow generated by radiative energy loss
is always smaller than observed data,
when we fit the nuclear modification factor.

In this proceeding, we consider jet-fluid string (JFS) formation
and its decay processes,
where a propagating jet parton picks up a partner parton
from the fluid to form a string. 
If we take account of JFS formation at the end of QGP evolution,
JFSs would have a large momentum anisotropy
coming from both of the hydrodynamically evolved fluid parton
and the jet parton which suffers the energy loss.
Furthermore, we can take account of both quarks and gluons
on the same footing in JFS formation.
We show the basic concept of the JFS model in Fig.\ref{Fig:JFSform}.
We find that JFS decay is an efficient process to produce high-$p_T$ hadrons,
and a large energy loss is required to describe the high-$p_T$ hadron spectra.
As a result of this large energy loss,
high-$p_T$ hadron elliptic flow becomes larger
than in the independent fragmentation (IF) picture,
while we slightly underestimate the observed $v_2$ of pions at high-$p_T$.

There are several attempts to include hadron formation processes
from jet (or shower) and fluid (or soft/thermal) partons.
Recombination processes of
thermal-thermal (TT),
thermal-shower (TS),
and shower-shower (SS) partons
have been considered to form hadrons,
which are considered to dominate
in a wide range of $p_T$ in Ref.\cite{Oregon}.
In the quark coalescence model,\cite{Texas}
coalescence of a soft parton and a quenched jet parton
(soft-hard coalescence)
is found to be important in intermediate-$p_T$ ($3<p_T<6$ GeV/$c$)
hadron production.
In a recombination model,\cite{Duke_New}
several processes to combine soft and hard partons are investigated.
In these models,
quarks are considered to form hadrons (including resonances) directly.
On the other hand,
we consider {\em gluons} as well as quarks can form {\em strings},
which decay into several hadrons in this work.

\section{JFS Model}

In the JFS model,
we have four ingredients to describe high-$p_T$ hadron distribution;
mini-jet production,
jet parton evolution in the QGP fluid,
jet-fluid string formation,
and its decay.
First,
mini-jet partons are generated in the pQCD framework
by using the \textsc{pythia} program (version 6.4).\cite{PYTHIA}
We have tuned \textsc{pythia} parameters,
$K$-factor ($K\sim 1.85$)
and minimum $p_T$ of jet partons ($p_{{\scriptscriptstyle T}0} \sim 2$ GeV),
to reproduce the high-$p_T$ pion spectrum in $p+p$ collisions
at RHIC.\cite{pp}
Secondly,
the energy loss of generated jet partons in the QGP fluid
is evaluated by using
the first order Gyulassy-L\'evai-Vitev (GLV) energy loss formula\cite{e-loss}
with simplification\cite{HN2004}:
\begin{equation}
\Delta E=C\times 3\pi\alpha_s^3 F_{\rm color}\int_{\tau_0}^\infty
d\tau \;\rho(\tau,\bm x(\tau)) \cdot (\tau-\tau_0)
\log\biggl(\frac{2E_0}{\mu^2L}\biggr) ,
\label{Eq:GLV}
\end{equation}
where 
$F_\text{color}$ is a color factor (1 and 9/4 for $q(\bar{q})$ and $g$),
$E_0$ is the initial energy,
$\alpha_s=0.3$, $\mu=0.5$ GeV, and $L=3$ fm.
We utilize the fluid evolution obtained by the 3D hydrodynamical model
calculation\cite{HN2004,HN2003}
with the Glauber type initial condition\cite{HT2002}
to evaluate parton density $\rho(\tau,\bm x)$,
which can be obtained on the web.\cite{HiranoWWW}
The overall energy loss factor $C$ is regarded
as an adjustable parameter
to reproduce $R_{AA}$,
and fixed as $C \sim 6$ by fitting pion $p_T$-spectrum
in mid-central Au+Au collisions as discussed later.
Low-$p_T$ ($p_T < 2\ \mathrm{GeV}/c$) partons are considered to be 
absorbed in the QGP fluid.
Next,
a jet parton is assumed to pick up a fluid parton
at the boundary of QCD phase transition
to form a color singlet string.
We here only consider a string formation ($\bar{q}q$ or $gg$)
whose mass is larger than 2 GeV.
Fluid parton momenta are obtained from Lorentz-boosted
Fermi (Bose) distribution for $q$ or $\bar{q}$ ($g$).
Light flavored ($u, d$ and $s$) fluid quarks
are regarded as massless
and the flavor is chosen with the same probability of $1/3$.
Effects of hadronic resonance formations for a mass smaller than 2 GeV
will be reported in the future.
Finally, the decay of these strings
producing multi-hadrons are evaluated
in the Lund string fragmentation model \textsc{pythia}.
We also show the results in the independent fragmentation (IF)
for comparison.

\section{Results and Discussions}

\begin{figure}[t]
\begin{center}
\includegraphics[width=6cm,clip]{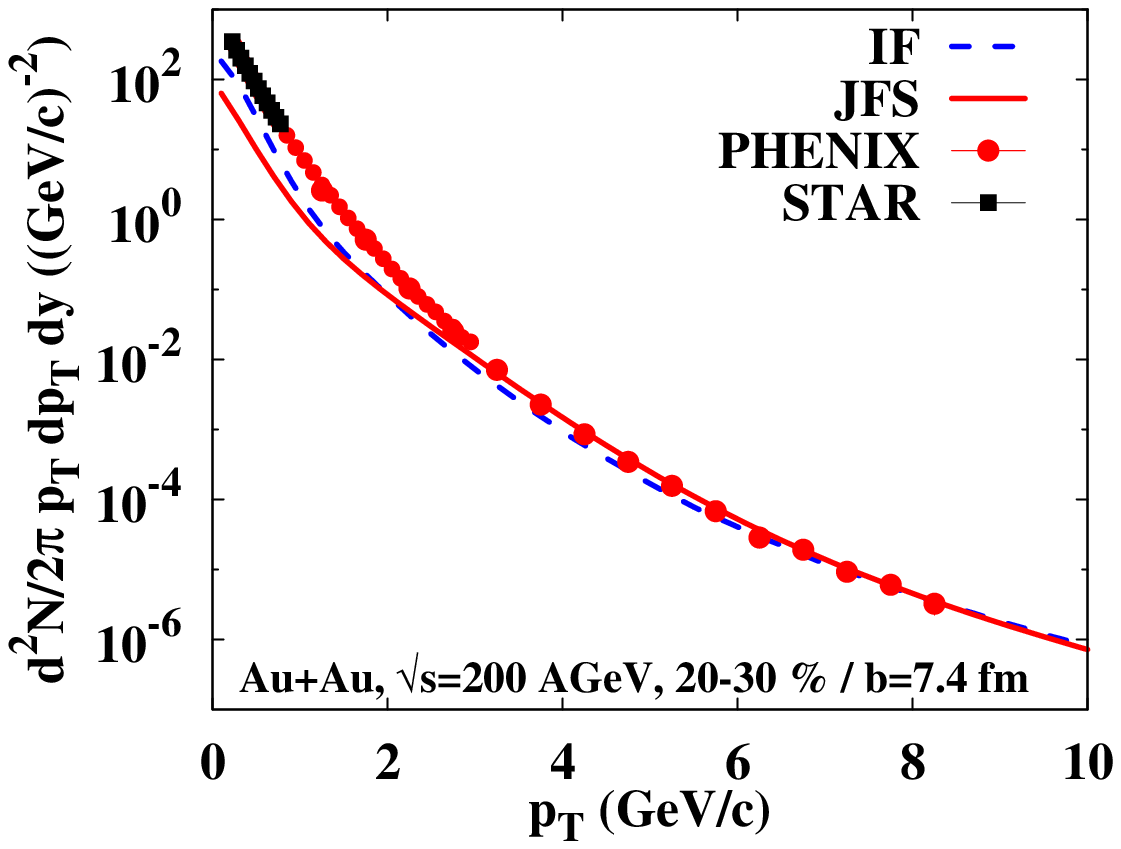}~
\includegraphics[width=6cm,clip]{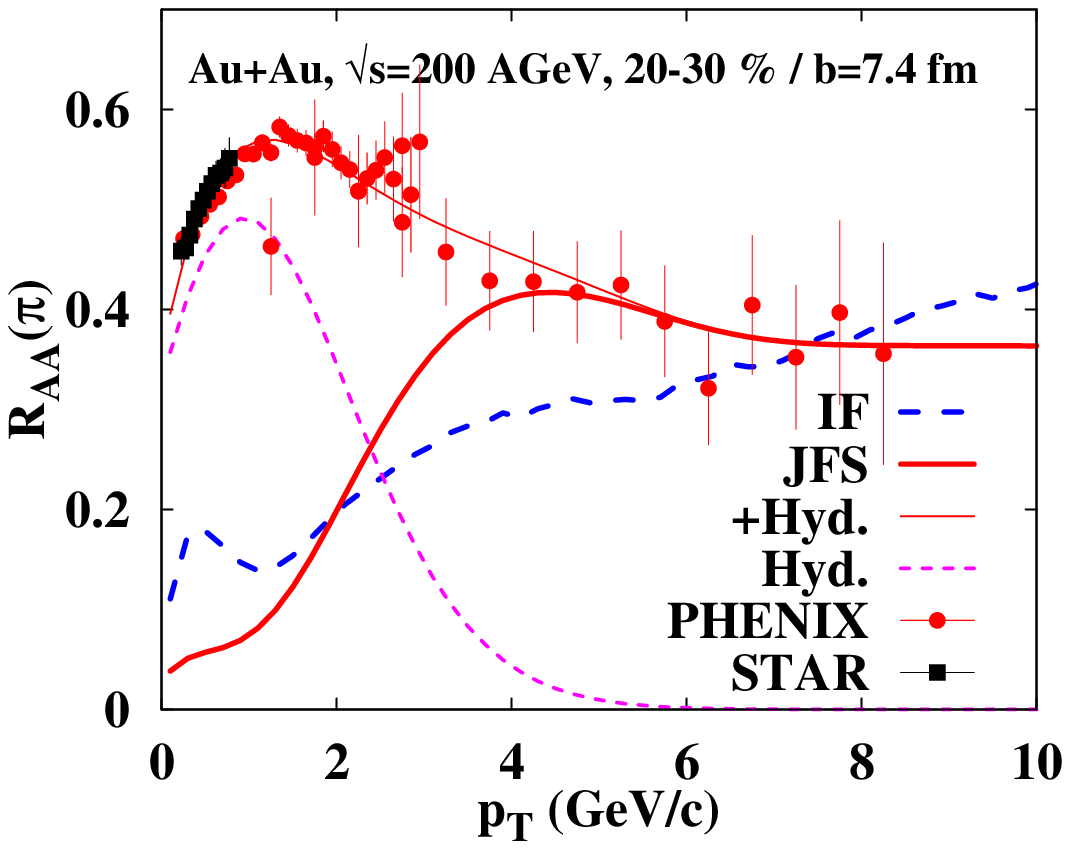}
\caption{
Left panel:
Charged pion $p_T$-spectrum in mid-central ($b=7.4$ fm) Au+Au collisions.
We show JFS results and those
combined with hydrodynamic exponential component.
Experimental data of $\pi^\pm$ are taken from STAR\protect\cite{pipstar}
and PHENIX\protect\cite{pipphen}, and $\pi^0$
multiplied by two are taken from PHENIX\protect\cite{pi0phen} data,
with corresponding centrality 20-30\%.
Right panel:
Two types of fragmentation scheme, JFS and IF,
are compared 
in $R_{AA}$ for mid-central collisions.
We see saturating behavior at high-$p_T$ in JFS,
and agrees with the data well.
}
\label{Fig:JFS-pt}
\end{center}
\end{figure}

In the left panel of Fig.~\ref{Fig:JFS-pt}, 
we show the calculated $p_T$-spectrum of charged pions
in mid-central Au+Au collisions at RHIC.
The energy loss factor of $C \sim 6$ is found to fit the high-$p_T$
part of the spectrum.
When JFS results are combined with the low momentum hydro dominant part
parametrized in the exponential form
$Ed^3N_{\rm Hydro}/dp^3=A_1\exp(-p_T/T_1)+A_2\exp(-p_T/T_2)$,
the calculated spectrum agrees
with experimental 
data from STAR\cite{pipstar} 
and PHENIX.\cite{pi0phen,pipphen}

It would be instructive to compare the JFS results
with those in a standard picture,
in which high-$p_T$ hadrons are formed 
through the IF of jet partons.\cite{HN2004}
In the right panel of Fig.~\ref{Fig:JFS-pt},
we compare the nuclear modification factor $R_{AA}$
in different hadronization schemes (JFS and IF)
by using the same fluid evolution profile.
Experimental values are pion
$p_T$-spectrum of Au+Au \cite{pipstar,pi0phen,pipphen}
normalized by that of $p+p$\cite{pp} and $N_{\rm coll}$.
The latter model (IF)
is similar to the Hydro+Jet model by Hirano and Nara\cite{HN2004}
at high-$p_T$,
but we have not taken care of the Cronin and shadowing effects,
which enhance the intermediate-$p_T$ hadrons.

In JFS and IF models,
different values of energy loss factor $C$ are needed
to fit high-$p_T$ spectrum,
$C_{\rm JFS}\sim 6$ and $C_{\rm IF}\sim 2$.
This is because high-$p_T$ hadron production is easier in JFS decay;
when a high-$p_T$ jet parton picks up
an approximately collinear low-$p_T$ fluid parton,
a light mass string is formed and decays into a fewer hadrons.
It is also due to the fragmentation scheme itself.\cite{NVC}
We also find a different $p_T$ dependence of $R_{AA}$.
While $R_{AA}$ saturates at around $p_T \sim 6\ \mathrm{GeV/c}$ in JFS, 
it slowly grows in the IF model.
The latter dependence may be compensated at intermediate-$p_T$
by the Cronin and shadowing effects,
but the high-$p_T$ dependence would remain.

\begin{figure}[t]
\begin{center}
\includegraphics[width=6cm,clip]{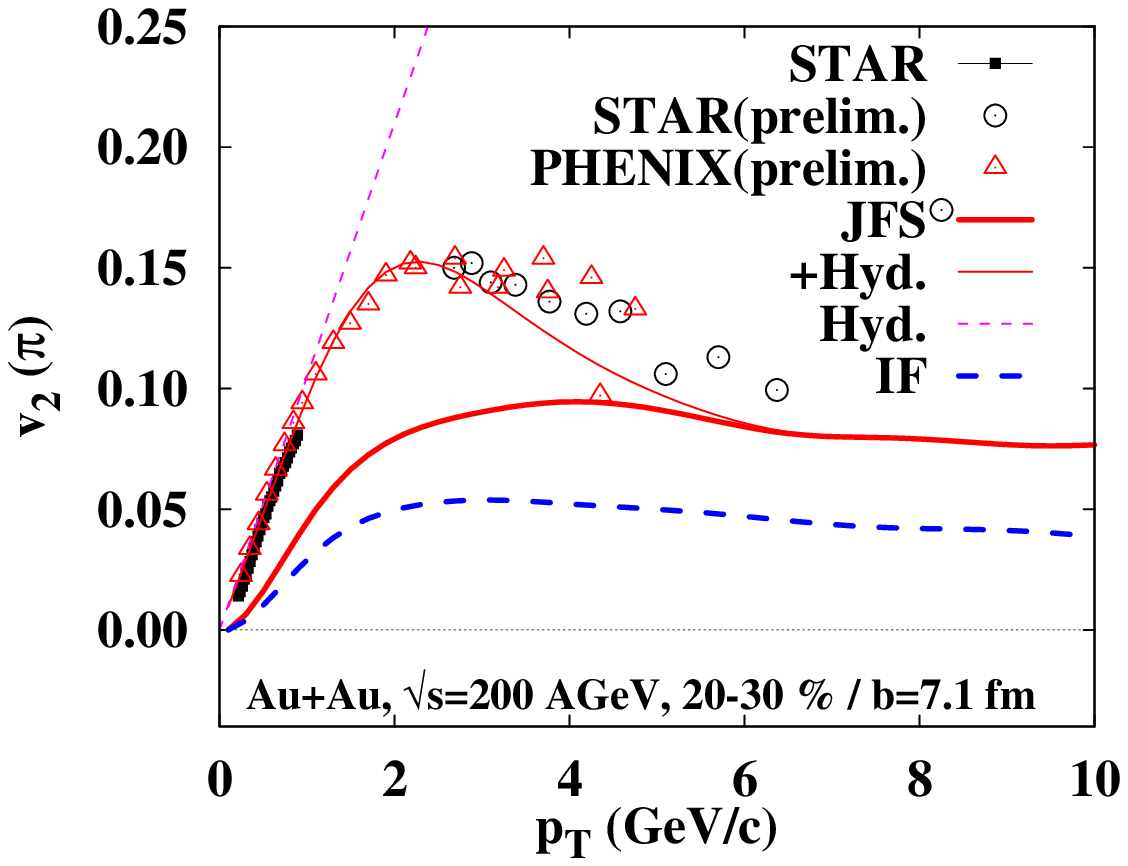}~%
\includegraphics[width=6.1cm,clip]{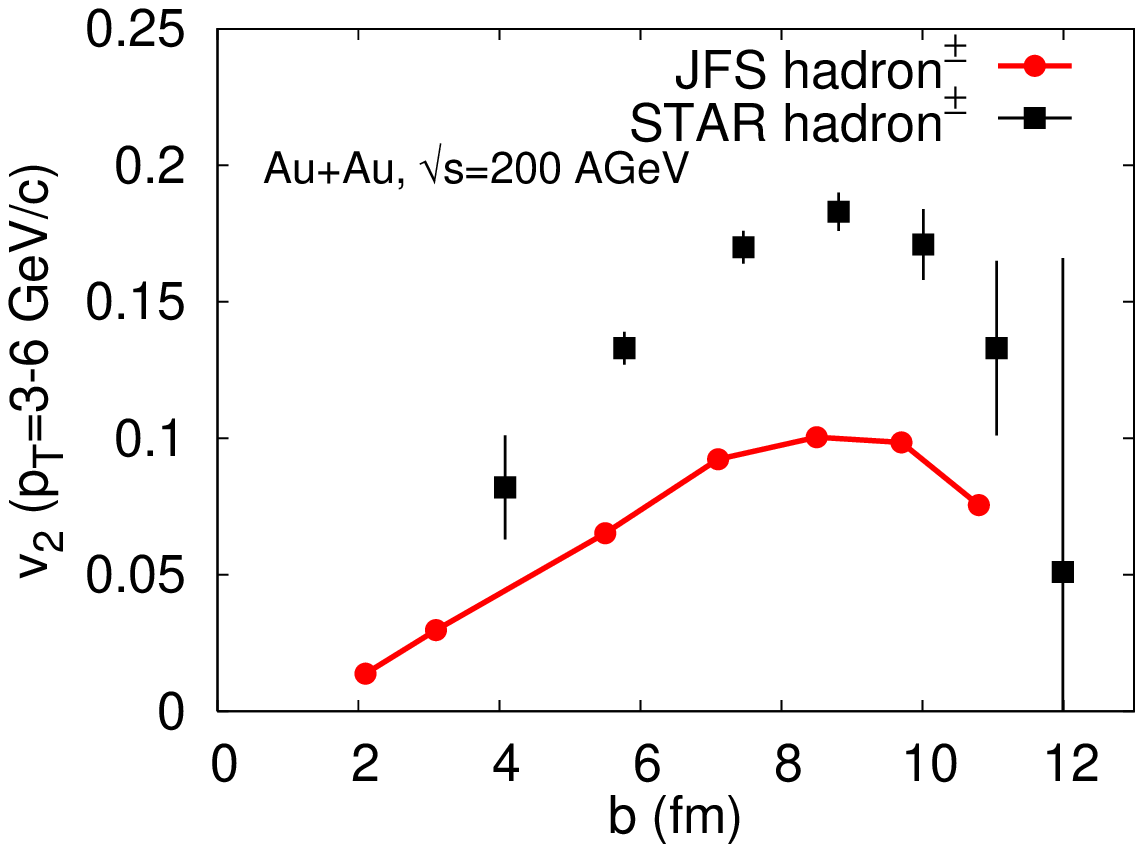}  
\caption{
Left panel:
Charged pion $v_2$ as a function of $p_T$.
We compare JFS results with 
STAR data\protect\cite{STARv2} of corresponding centrality 20-30\%,
and PHENIX and STAR preliminary data\protect\cite{Sorensen} of 
minimum-bias events.
Right panel:
Integrated $v_2$ of charged hadrons in the range of $3<p_T<6$ GeV/c
versus impact parameter $b$.
We compare
JFS results are compared with STAR data\protect\cite{STARv2}.
}
\label{Fig:v2}
\end{center}
\end{figure}

Next, we evaluate the elliptic flow $v_2$ in the JFS model.
In the left panel of Fig.~\ref{Fig:v2},
we show the calculated $p_T$ dependence of charged pion $v_2$
in comparison with the experimental data
of charged pions.\cite{STARv2b,Sorensen}
We find that the JFS results are around two times larger than IF results,
and are close to the data at high-$p_T$;
the data may be around 0.10, and JFS gives $\sim$ 0.08 at $p_T > 6$ GeV/$c$.
The difference between JFS and IF mainly comes from
the energy loss strength (parameter $C$ in Eq.~(\ref{Eq:GLV})),
and the picked fluid parton $v_2$ also contributes
to enhance the hadron $v_2$ by around 0.01.
The high-$p_T$ $v_2$ values in JFS is comparable with
the results in the Recombination+Fragmentation model\cite{Duke}
($v_2 \sim 0.10$ at $p_T\sim 5-10$ GeV/c),
and Hydro+Jet model\cite{HN2004}
($v_2 \sim 0.10$ at $p_T\sim 3$ GeV/c).
In the former,
the parton energy loss is parameterized by the angle dependent path length,
then a sharp edge density distribution is implicitly assumed.
In the latter,
$v_2$ decreases and underestimates the data
at higher $p_T$ ($p_T > 5\ \mathrm{GeV}/c$).
We also show the combined results of JFS and the Hydro component,
whose relative weight is already fixed from the $p_T$-spectrum fit.
Combined results well explain $v_2$ data up to around 2 GeV/$c$
with Hydro component having $v_2 \propto p_T$.
We clearly find that we underestimate the data at intermediate-$p_T$
($3-6$ GeV/$c$), where recombination processes would be important.

In the right panel of Fig.\ref{Fig:v2},
we plot the impact parameter dependence of
charged hadron $v_2$ integrated in the range of $3<p_T<6\ \mathrm{GeV}/c$
in comparison with STAR data\cite{STARv2}
obtained by four-particle cumulant method.
We note that the calculated $v_2$ values are only around half of the data.
However, we would like to point out that
the present $v_2$ values are
two times larger than those in a simple simulation
with Woods-Saxon density distribution\cite{STARv2}
and comparable to those in a simple calculation
with hard-sphere density profile (maximum $v_2\sim 0.10$),\cite{STARv2}
while we simulate the parton dynamics
with time-evolving 3D hydrodynamics.

\section{Summary}
We have proposed a Jet-Fluid String (JFS) model
as a mechanism to produce high-$p_T$ hadrons.
In the JFS model, following components are combined;
mini-jet production in pQCD,
energy loss with simplified GLV formula,
3D hydrodynamic simulations,
and Lund string decay.
JFS decay is found to produce high-$p_T$ hadrons effectively,
and we can utilize the 
3D hydrodynamical expansion.
After fitting high-$p_T$ spectrum in Au+Au collisions,
we find that the calculated $v_2$ values roughly reproduce
the data at high-$p_T$.
At intermediate-$p_T$,
JFS results of $v_2$ are about two times larger than in the independent
fragmentation, but still they are around half of the data.
This underestimate may be due to the lack of lower mass ($<2$ GeV)
string or resonance, or the Fluid-Fluid String formation.
From these results, we conclude that
JFS would be a plausible fragmentation scheme to produce high-$p_T$ hadrons
in relativistic heavy-ion collisions. 

One of the authors (MI) is grateful to M.~Asakawa
for fruitful discussions. 
This work was supported in part
by the 21st Century COE Program 
``Towards a New Basic Science; Depth and Synthesis'',
Osaka University (MI),
and 
by the Ministry of Education,
Science, Sports and Culture, Grant-in-Aid for Scientific Research
under the grant numbers,
    18-10104 (TH),		
    15540243 (AO),		
and 
    1707005 (AO).		

\end{document}